\documentstyle[12pt]{article}
\begin{document}
\begin{center}
\textbf{MAGNETOHYDRODYNAMICS  IN  THE  INFLATIONARY  UNIVERSE}\\
\bigskip
\bigskip
\bigskip
I. Brevik\footnote{E-mail address:  iver.h.brevik@mtf.ntnu.no} and H. B. Sandvik\footnote{E-mail address:  h.sandvik@ic.ac.uk}\\
\bigskip
Division of Applied Mechanics,\\
 Norwegian University of Science and Technology,\\
N-7491 Trondheim, Norway
\bigskip
Revised version, December 1999\\
\end{center}
\bigskip
\bigskip
\begin{abstract}
Magnetohydrodynamic (MHD) waves are analysed in the early Universe, in the inflationary era, assuming the Universe to be  filled with  a nonviscous fluid of the Zel'dovich type ($p=\rho$) in a metric of the de Sitter form. A spatially uniform, time dependent, magnetic field ${\bf B_0}$ is assumed to be present. The Einstein equations are first solved to give the time dependence of the scale factor, assuming that the matter density, but not the magnetic field, contribute as source terms. The various  modes are thereafter analysed; they turn out to be essentially of the same kind as those encountered in conventional nongravitational MHD, although the longitudinal magnetosonic wave is not interpretable as a physical energy-transporting wave as the group velocity becomes superluminal.  We determine the phase speed of the various modes; they turn out to be  scale factor independent. The Alfv\'{e}n velocity of the transverse magnetohydrodynamic wave becomes extremely small in the inflationary era, showing that the wave is in practice 'frozen in'. 
\end{abstract}

\newpage
\section{Introduction}

There exists the possibility that a primordial magnetic field was created at some early stage in the early Universe. This topic has attracted considerable attention in the recent past. The important point is that such a field may have left an observable imprint in galaxies. Our Galaxy, as well as many other spiral galaxies, is endowed with coherent magnetic fields, ordered on scales larger than about 10 kpc, with a typical strength of   $3 \, \mu$G (= $3 \times 10^{-6}$ gauss) \cite{sofue86}, \cite{turner88}. This corresponds to a magnetic field energy density $B^2/8\pi$ which is of the same order of magnitude as the observed energy density $\rho_{\gamma}$ of the microwave background: $\rho_{\gamma} \sim (4\, \mu G)^2/8\pi $ \cite{olinto98}. 

There are two plausible explanations for the existence of these large scale galactic fields. The most popular explanation is that such fields are the result of  a dynamo amplification \cite{landau84} of a weak seed field \cite{kronberg94}, created some time in the early Universe, for instance at the electroweak phase transition \cite{baym96}, \cite{son99}. This transition took place at the instant $t \sim 10^{-10}$ s, corresponding to a temperature of $T_c \sim 10^{15}$ K   (100 GeV). The size of the event horizon was at this instant  $H_{ew}^{-1} \sim 10$ cm.  If the galactic dynamo is efficient at amplification, one estimates  the magnetic seed field $B_{seed}$ to be lying in the range between $10^{-23}$ G and $10^{-19}$ G  \cite{olinto98}, \cite{brandenberger99}. Another plausible explanation  for the existence of the present magnetic fields is that they could have originated from a relatively large primordial seed field amplified by the collapse of a galaxy \cite{olinto98}.

Generally speaking, the magnetohydrodynamic theory (MHD) in curved spacetime is a relatively new development in astrophysics. We may mention, therefore, that useful recent review articles are given by Olinto \cite{olinto98} and Enqvist \cite{enqvist97}.

Whereas in previous works it has often been assumed that the seed magnetic field is created in the electroweak transition region, we will in the present paper go further back in time and consider instead the {\it inflationary era}. As is commonly assumed, this era took place from $t \sim 10^{-35}$ s to $t \sim 10^{-33}$ s.  The Universe was then in a state of violent expansion, with a cosmological scale factor being determined essentially from the cosmological constant $\Lambda $. If the influence from matter was negligible, as one often assumes, the metric was a pure de Sitter metric with scale factor proportional to $\exp {(\sqrt{\frac{\Lambda}{3}}t}) $. During the inflationary period the Universe was subject to a large increase in size - one often assumes an increase of the order $10^{50}$ - but the expansion might actually have been much larger. The temperature correspondingly fell from $T \sim 10^{27}$ K to $T \sim 10^{22}$ K. The Universe consisted during this era of a vacuum "fluid" with energy density $ \rho_{vac} = \Lambda /8 \pi G $, and extreme tensile stress, $ p_{vac}= -\Lambda /8 \pi G$ (this negative pressure giving rise to the repulsive gravitation). 

Brandenburg {\it et al.} \cite{brandenburg96} investigated, on the basis of a flat FRW fluid model, whether a primordial magnetic field decreases or increases with time. This work was related to prior work of Gailis {\it et al.} \cite{gailis95}. In the following, we will instead assume that there was a fluid of the Zel'dovich type in the early Universe.  This means that the equation of state of the fluid was $p=\rho$, corresponding to a velocity of sound being equal to the velocity of light. There are reasons to expect that this extreme kind of fluid was present in the beginning, and at the end, of the inflationary era \cite{gron90}. Here, both $\rho$ and $p$ are taken to act as sources in the Einstein equations.  In addition, we assume that there was a primordial magnetic field ${\bf B_0}$ present. Since the magnitude of ${\bf B_0}$ was so small, the magnetic field energy was too small to have any appreciable influence on the metric. We therefore examine simply MHD effects on a fixed metric, the metric being determined by $\Lambda, \rho $, and $ p $. Note that because of the matter, the scale factor does no longer have the simple form $\exp{( \sqrt{\frac{\Lambda}{ 3}}t})$ mentioned above.

Our main purpose in the following is to establish the governing equations of the combined system, matter plus field. As far as we know, this has not been done before. We will establish the plane, linear, MHD wave modes on the spatially uniform de Sitter background. A characteristic property of this kind of system is that $\rho R^6 = $ const. ($R$ being the scale factor), instead of the conventional relationship $\rho R^4$=const. found in the radiation dominated FRW case.  If $ {\bf B_0}$ is perpendicular to ${\bf k}$, we find that there is a longitudinal magnetosonic wave whose phase velocity is superluminal. Since this wave is nondispersive, its group velocity becomes accordingly also superluminal, so that this wave is not physically acceptable as an energy-transporting wave. Next, if ${\bf B_0}$ is parallel to ${\bf k}$, there is a simple luminal pressure wave, and there is also a transverse magnetohydrodynamic wave propagating with the characteristic de Sitter variant of the Alfv\'{e}n velocity  given by Eq. (50) below. In the inflationary era, the magnitude of the Alfv\'{e}n velocity is extremely small.

Numerical investigations of the time development of a primordial magnetic field, along the lines of \cite{brandenburg96}, will not be undertaken in this paper.

We adopt in the following Heaviside - Lorentz, instead of Gaussian, electromagnetic units.

\section{The de Sitter - Zel'dovich scale factor}

We use the convention in which the Minkowski metric is $\eta_{\mu\nu}=(-+++)$, Greek indices are summed from 0 to 3, Latin indices are summed from 1 to 3. We let $U^{\mu}=(U^0, U^i)$ designate the four-velocity of the cosmic fluid.

When the cosmological constant $\Lambda (>0)$ is included, it is convenient to write Einstein's equations on the form
\begin{equation}
G_{\mu\nu} \equiv R_{\mu\nu}-\frac{1}{2}Rg_{\mu\nu}=8\pi G \tilde{T}_{\mu\nu},
\end{equation}
\label{1}
where $\tilde{T}_{\mu\nu}$ is the modified energy-momentum tensor
\begin{equation}
\tilde{T}_{\mu\nu}=T_{\mu\nu}-\frac{\Lambda}{8\pi G}g_{\mu\nu},
\end{equation}
\label{2}
with $T_{\mu\nu}=(\rho+p)U_\mu U_\nu +pg_{\mu\nu}$ being the ordinary energy-momentum tensor of an ideal (non-viscous) fluid. Defining the modified energy density $\tilde \rho$ and pressure $\tilde p$ by
\begin{equation}
\tilde\rho= \rho+\frac{\Lambda}{8\pi G}, ~~~~~\tilde p= p-\frac{\Lambda}{8\pi G},
\end{equation}
\label{3}
we can write (2) as
\begin{equation}
\tilde{T}_{\mu\nu}=(\tilde\rho+\tilde p)U_{\mu}U_{\nu}+\tilde p g_{\mu\nu}.
\end{equation}
\label{4}
Consider now the line element having the de Sitter form:
\begin{equation}
ds^2=-dt^2+R^2(t)[ dr^2+r^2(d\theta^2+\sin^2\theta d\varphi^2)].
\end{equation}
\label{5}
This form refers to comoving coordinates ( $r$ is constant for a fixed matter element). We consider the Einstein equations in the orthonormal frame (designated by "hats"). We need the following components of the Einstein tensor:
\begin{equation}
G_{\hat 0}^{\hat 0}=-\frac{3 \dot R^2}{R^2}, ~~~~~G_{\hat r}^{\hat r}=-\frac{2 \ddot R}{R}
-\frac{\dot R^2}{R^2}.
\end{equation}
\label{6}
The "energy" equation $G_{\hat 0}^{\hat 0}=8\pi G \tilde {T}_{\hat 0}^{\hat 0}$ now becomes
\begin{equation}
\frac{\dot R^2}{R^2}=\frac{8\pi G}{3}\tilde \rho,
\end{equation}
\label{7}
whereas the "pressure" equation $G_{\hat r}^{\hat r}=8\pi G \tilde{T}_{\hat r}^{\hat r}$ becomes
\begin{equation}
\frac{2\ddot R}{R}+\frac{\dot R^2}{R^2}=-8\pi G \tilde p.
\end{equation}
\label{8}
These are the field equations. We have also the energy conservation equation which for the de Sitter - Zel'dovich fluid leads to the behaviour $\rho \propto R^{-6}$, instead of the behaviour $\rho \propto R^{-4}$ that is characteristic for the radiation dominated era.

Let us now determine $R(t)$, by first making the substitution
\begin{equation}
\frac{\rho}{\rho_0}=\left( \frac{R}{R_0} \right)^{-6}.
\end{equation}
\label{9}
Here $R_0$ and $\rho_0$ refer to an initial instant $t=t_0$, which marks the beginning of the inflationary period. The first order equation (7) becomes
\begin{equation}
\frac{\dot R^2}{R_0^2}=\frac{8\pi G \rho_0}{3}\left( \frac{R}{R_0} \right) ^{-4}+ \frac{\Lambda}{3}
\left( \frac{R}{R_0}\right)^2.
\end{equation}
\label{10}
Solving this equation for $dt $ and integrating from $t=t_0$ to $t$, corresponding to an increase of the scale factor from $R_0$ to $R$, we get
\begin{equation}
\frac{R(t)}{R_0}=\left[ e^{\sqrt{3\Lambda}(t-t_0)}+\frac{8\pi G\rho_0/\Lambda}
{1+\sqrt{1+8\pi G\rho_0/\Lambda}}\sinh [\sqrt{3\Lambda}(t-t_0)] \right]^{\frac{1}{3}}.
\end{equation}
\label{11}
This expression gives the de Sitter - Zel'dovich scale factor. At first sight it may appear as if the Zel'dovich state equation $p=\rho$ plays no role in  Eq. (11) at all, since in the derivation of this equation we  integrated the "energy" equation (7) which in itself makes no reference to the fluid's state equation. However, in the derivation  we made explicit use of Eq. (9), which is characteristic for a Zel'dovich fluid. Therefore, Eq. (11) implicitly relies upon the Zel'dovich property.  

We see that for empty space ($\rho_0=0 $) the expression reduces to $R(t)/R_0 = \exp [\sqrt{\frac{\Lambda}{3}}(t-t_0)]$, which is the conventional de Sitter form. For $t=t_0$, $R(t_0)=R_0$, as it should.

\section{Magnetohydrodynamics in the early universe}

\subsection{Maxwell's equations}

As an introductory step, before embarking upon the energy-momentum conservation problem it is instructive to consider Maxwell's equations themselves, formulated in a curvilinear basis which we will choose in the present subsection to be the spherical basis.

Let the spatial coordinates be numerated as $x^i = (r, \theta, \varphi)$, corresponding to the basis vectors $( {\bf e}_r, {\bf e}_{\theta}, {\bf e}_{\varphi})$.  The determinant of the spatial metric is, according to (5),
\begin{equation}
\gamma = \rm{det} (g_{ij})=R^6 r^4 \sin^2 \theta.
\end{equation}
\label{12}
We introduce the electromagnetic field tensor \cite{moller72}, \cite{landau75}, \cite{bladel84}, \cite{brevik87}
\begin{eqnarray}
F_{\mu\nu}= \left( \begin{array}{cccc}
0		&		-E_r			&	-E_{\theta}			&	-E_{\varphi}\\
E_r		&      		0		&	\sqrt{\gamma}B^{\varphi}&	-\sqrt{\gamma}B^{\theta}\\ 
E_{\theta}  & -\sqrt{\gamma}B^{\varphi}   &	0				&	\sqrt{\gamma}B^r\\
E_{\varphi} &	\sqrt{\gamma}B^{\theta} &	-\sqrt{\gamma}B^r		&      0
\end{array}
\right)
\end{eqnarray}
\label{13}
and the dual tensor density ${\cal{F}}^{\mu\nu}=\sqrt{-g} F^{\mu\nu}$:
\begin{eqnarray}
\cal{F}^{\mu\nu}=\left( \begin{array}{cccc}
0				 &	\sqrt{\gamma}E^r		&	\sqrt{\gamma}E^{\theta}	 &	\sqrt{\gamma}E^{\varphi}\\
-\sqrt{\gamma}E^r	       &	0				&      B_{\varphi}		 &	-B_{\theta}\\
-\sqrt{\gamma}E^{\theta} &	-B_{\varphi}		&	0				 &	 B_r \\
-\sqrt{\gamma}E^{\varphi}&     B_{\theta} 		&	-B_r				 &      0 
\end{array}
\right).
\end{eqnarray}
\label{14}
Three-dimensional indices are raised and lowered by means of the antisymmetric pseudotensor of rank 3:
\begin{equation}
\epsilon_{ijk}=\gamma^{1/2}\delta_{ijk},~~~~~\epsilon^{ijk}=\gamma^{-1/2}\delta_{ijk},
\end{equation}
\label{15}
$\delta_{ijk}$ being the antisymmetric Levi-Civita symbol with $\delta_{123}=1$. As $g_{00}=-1$, the gravitational permittivity and permeability are $\varepsilon = \mu = (-g_{00})^{-1/2}=1$, and we do not have to distinguish between the electric field ${\bf E}$ and the electric induction ${\bf D}$, nor between the magnetic field ${\bf H}$ and the magnetic induction ${\bf B}$.

With $J^{\mu}=(\rho, {\bf J})$ being the electromagnetic four-current density, the four-dimensional Maxwell equations
\begin{equation}
F_{[\mu\nu,\rho]}=0, ~~~~~(-g)^{-1/2} {\cal{F}^{\mu\nu}}_{,\nu}= J^{\mu}
\end{equation}
\label{16}		
can be expressed on three-dimensional form as
\begin{equation}
({\bf\nabla\times \bf E})^i = -\frac{1}{\sqrt{\gamma}}\partial_0(\sqrt{\gamma} B^i),
~~~~~{\bf\nabla}\cdot{\bf B}=0,
\end{equation}
\label{17}
\begin{equation}
({\bf\nabla \times\bf B})^i= J^i+\frac{1}{\sqrt{\gamma}}\partial_0(\sqrt{\gamma}\,E^i),
~~~~~{\bf\nabla}\cdot{\bf E}=\rho_e,
\end{equation}
\label{18}
the curl and divergence operators being defined as
\begin{equation}
({\bf\nabla \times \bf E})^i=\epsilon^{ijk}\partial_jE_k,
~~~~~{\bf\nabla}\cdot {\bf B}=\gamma^{-1/2}\partial_i
(\sqrt{\gamma}\,B^i).
\end{equation}
\label{19}

\subsection{Energy-momentum balance equations}

After having given above the general scheme for Maxwell's equations in curvilinear coordinates, we shall in the following simplify the formalism by using "Cartesian" coordinates $x,y,z$, whereby the position four-vector becomes expressible as $x^\mu = (t,x,y,z)$. The line element is written as 
\begin{equation}
ds^2=-dt^2+R^2(t)(dx^2+dy^2+dz^2).
\end{equation}
\label{20}
The Christoffel symbols become simple: $\Gamma_{ij}^0 = R \dot{R}\delta_{ij}$,  $\Gamma_{0j}^i=\Gamma_{j0}^i=(\dot{R}/R)\delta_{ij}$. Since $g_{00}=-1$ and $g_{0i}=0$ the Lorentz factor takes the same form as in special relativity (cf. Sect. 10.3 in \cite{moller72}):
\begin{equation}
\gamma=(1-u^2)^{-1/2},
\end{equation}
\label{21}
where $u^i = dx^i/dt $ are the curvilinear components of the three-dimensional coordinate velocity ${\bf u}$ whose square is $u^2=g_{ij}u^i u^j$. The four-velocity of the fluid is $U^{\mu}=\gamma (1, u^i)$. The energy-momentum balance equations for the mechanical subsystem are $T_{\mu ; \nu}^{\nu}=f_{\mu}$, where $T^{\mu\nu}$ is the energy-momentum tensor of the fluid as given above, and $f_{\mu}$ is the electromagnetic four-force density:
\begin{equation}
f_{\mu}=F_{\mu\nu} J^{\nu} = (-{\bf E \cdot J},\; \rho_e E_i + ({\bf J\times B})_i),
\end{equation}
\label{22}
the vector product in  curvilinear coordinates being defined as
\begin{equation}
({\bf J\times B})_i=\epsilon_{ijk}J^j B^k.
\end{equation}
\label{23}
The energy-momentum balance equations for the fluid mechanical subsystem are
\begin{equation}
\frac{1}{\sqrt{-g}}\partial_\nu(\sqrt{-g}T_\mu^{\nu})-\frac{1}{2}(\partial_{\mu}g_{\alpha\beta})T^{\alpha\beta}
=f_{\mu}.
\end{equation}
\label{24}
We assume the fluid to be an ideal MHD fluid, implying that the electrical conductivity $\sigma \rightarrow \infty$. It corresponds to a zero electric field ${\bf E}' =0$ in a frame of reference moving with the fluid volume element considered. As ${\bf J}$ is finite, we then have that ${\bf E}= -{\bf u\times B}$, and  the volume force density ${\bf f}$ becomes equal to ${\bf J\times B}$. For $\mu=0$ we get from (24), with the definition $ {\bf S}=(\rho+p)\gamma^2 {\bf u}$,
\begin{equation}
\frac{1}{R^3}\partial_0[ R^3(\rho+p)\gamma^2]-\partial_0 p+{\bf \nabla \cdot S}
+\frac{\dot R}{R}\,{\bf S \cdot u}= {\bf E \cdot J}.
\end{equation}
\label{25}
For $\mu =i$ we obtain analogously from (24)
\begin{equation}
\frac{1}{R^3}\partial_0(R^3S_i)+\partial_k(S_i u^k)+\partial _i p=({\bf J\times B})_i.
\end{equation}
\label{26}
Consider next the electromagnetic subsystem. In the present coordinate system the field tensor schemes analogous to (13) and (14) are
\begin{eqnarray}
F_{\mu\nu}=\left( \begin{array}{cccc}
0                &        -E_1          &          -E_2        &        -E_3          \\
E_1              &          0           &           R^3B^3     &       -R^3B^2        \\
E_2              &         -R^3B^3      &            0         &         R^3B^1       \\
E_3              &          R^3B^2      &           -R^3B^1    &          0
\end{array}
\right),
\end{eqnarray}
\label{27}
\begin{eqnarray}
\cal{F}^{\mu\nu}=\left( \begin {array}{cccc}
0                  &        R^3E^1       &           R^3E^2     &       R^3E^3        \\
-R^3E^1            &         0           &          B_3         &        -B_2         \\
-R^3E^2            &         -B_3        &           0          &         B_1         \\
-R^3E^3            &          B_2        &           -B_1       &          0       
\end{array}
\right) .
\end{eqnarray}
\label{28}
With $S^{\mu\nu}$ denoting the electromagnetic energy-momentum tensor the balance equations for this subsystem can be written as $-S_{\mu;\nu}^{\nu}=f_{\mu}$, where
\begin{equation}
S_\mu^{\nu}=F_{\mu\alpha}F^{\nu\alpha}-\frac{1}{4}g_{\mu}^{\nu}F_{\alpha\beta}F^{\alpha\beta}.
\end{equation}
\label{29}
In the present case where only the magnetic field contributes we have
\begin{equation}
f_i=(B^k\partial_k)B_i-\frac{1}{2}\partial_iB^2.
\end{equation}
\label{30} 
So far, the equation of state for the fluid has not been used. We now insert the Zel'dovich equation $p=\rho$, and assume small bulk velocities whereby $\gamma^2 \approx 1$. The term containing ${\bf S\cdot u}$ in the $\mu=0$ equation (25) is of order $u^2$ and is negligible. We obtain the approximate equation
\begin{equation}
\frac{1}{R^6}\partial_0(R^6\rho)+2{\bf \nabla \cdot(}\rho{\bf u)}={\bf E \cdot J}.
\end{equation}
\label{31}
The $\mu=i$ equation (26) becomes analogously
\begin{equation}
\frac{2}{R^3}\partial_0(R^3\rho u_i)+\partial_i \rho= (B^k\partial_k)B_i-\frac{1}{2}\partial_i B^2.
\end{equation}
\label{32}
When expressed on vector form,
\begin{equation}
\frac{2}{R^3}\partial_0(R^3\rho {\bf u})+{\bf \nabla}\rho={\bf (B\cdot \nabla)B}-\frac{1}{2}{\bf \nabla}B^2,
\end{equation}
\label{33}
this equation holds in an orthonormal basis as well.

Equations (31) and (32) are starting points for the perturbative theory below. In addition we shall need also the first of Maxwell's equations (17), written as a vector equation. Observing again that ${\bf E} = - {\bf u \times B}$, we have
\begin{equation}
{\bf \nabla \times(u\times B)}=\frac{1}{R^3}\partial_0(R^3 {\bf B}).
\end{equation}
\label{34}
It should be noted that when writing Maxwell's equations, we have not restricted  the magnetic  field to be  weak. As far as the magnetic field is concerned, the only restriction made on it is that it is not so strong that it contributes as a source term in Einstein's equations. Our remaining physical assumptions, to summarize, are the presence of an infinite electrical conductivity, nonrelativistic fluid velocities, and the Zel'dovich state equation.

\section{Magnetohydrodynamic waves}

We will now consider plane, linear MHD waves on the spatially uniform background in the inflationary universe. The background is specified by the magnetic field ${\bf B_0}$, fluid density $\rho_0$, and pressure $p_0$. These quantities are time dependent.  We shall need the proportionalities
\begin{equation}
\rho_0(t) \propto \frac{1}{R^6(t)}, ~~~~~B_0^i(t) \propto \frac{1}{R^3(t)}.
\end{equation}
\label{35}
The first of these proportionalities follows from Eq. (9). The second follows, in the zeroth order approximation, from Eq. (34) when both sides of this equation are expanded around the background. Generally, we expand as follows:
\begin{equation}
\rho=\rho_0+\delta\rho, ~~~~B^i= B_0^i+\delta B^i,
\end{equation}
\label{36}
with $\delta \rho$ and $\delta B^i$ being small quantities.

From Maxwell's equation $ {\bf \nabla \times B}={\bf J} + R^{-3}\partial_0(R^3 {\bf E})$ it follows that
\begin{equation}
{\bf J}= {\bf \nabla \times \delta B}+\frac{1}{R^3}\partial_0 [R^3({\bf u \times B_0})],
\end{equation}
\label{37}
which shows that ${\bf J}$ is a first order quantity. As ${\bf E}$ is also of first order, it is seen that the term ${\bf E \cdot J}$ on the right hand side of (31) is negligible. We obtain from (31), (32), and (34)
\begin{equation}
\frac{1}{R^6}\partial_0(R^6\delta\rho)+2\rho_0{\bf \nabla\cdot u}=0,
\end{equation}
\label{38}
\begin{equation}
2\rho_0\dot u_i+\frac{6 \dot R}{R}\rho_0u_i+\partial_i\delta\rho =
({\bf B_0 \cdot \nabla)}\delta{\bf B}_i -{\bf B}_0\cdot \partial_i\delta{\bf B},
\end{equation}
\label{39}
\begin{equation}
{\bf \nabla \times (u\times B}_0)=\frac{1}{R^3}\partial_0(R^3\delta {\bf B})
\end{equation}
\label{40}
(recall that we are using curvilinear coordinates for which the curl and vector product operators are as given by (19) and (23)).

We now expand the wave quantities as follows:
\begin{equation}
u^i=u_0^i e^{i\Theta}, ~~~\delta\rho=\frac{C}{R^6}e^{i\Theta},~~~\delta B^i=\frac{b_0^i}{R^3}e^{i\Theta},
\end{equation}
\label{41}
where $\Theta$ is the phase
\begin{equation}
\Theta=k_{\mu}x^{\mu}=k_ix^i+k_0t={\bf k\cdot r}-\omega t,
\end{equation}
\label{42}
and $u_0^i, C, b_0^i$ are constants. (We thus let the frequency $\omega$ refer to the {\it coordinate} time $t$, not to the {\it conformal} time.) Substitution into (38) and (39) yields
\begin{equation}
2\rho_0 R^6 {\bf k\cdot u_0}=C\omega,
\end{equation}
\label{43}
\begin{equation}
2\omega\rho_0 R^3\left( 1+3i\frac{\dot R}{R\omega}\right)u_{0i}-\frac{C}{R^3}k_i=
({\bf b_0\cdot B_0)}k_i-({\bf k\cdot B_0})b_{0i}.
\end{equation}
\label{44}
Since we are here considering waves moving on a "slowly" varying background (in complete analogy with  surface waves travelling on slowly varying currents or slowly varying topology in ordinary hydrodynamics), we can assume that $ \omega$ is much larger than the time derivative of the logarithmic scale factor:  $\omega \gg (d/dt) \ln R $.
The second term in the first bracket in (44) is thus negligible, and we get
\begin{equation}
2\omega\rho_0 R^3 u_{0i}-\frac{C}{R^3}k_i=({\bf b_0 \cdot B_0})k_i-({\bf k \cdot B_0})b_{0i}.
\end{equation}
\label{45}
The third governing equation follows from Maxwell's equation (40):
\begin{equation}
({\bf k \cdot B_0})u_0^i = ({\bf k \cdot u_0})B_0^i -\frac{\omega}{R^3}b_0^i.
\end{equation}
\label{46}
We now take the vector ${\bf k}$ to lie along the $x$ axis, and distinguish between two cases:

Let ${\bf B_0}$ be {\it perpendicular} to ${\bf k}$. From Eq. (45) it follows that $u_{02}=u_{03}=0$, so that the fluid velocity ${\bf u_0}$ is a purely longitudinal vector. From (46) it then follows that ${\bf b_0}$ is transverse, collinear with ${\bf B_0}$. The condition that the system determinant of the governing equations be equal to zero, yields the dispersion relation
\begin{equation}
\omega^2= \left( 1+\frac{{\bf B_0}^2}{2\rho_0} \right){\bf k}^2
\end{equation}
\label{47}
(${\bf k}^2=k^1 k_1$). From expression (42) for the phase it follows that the contravariant component of the longitudinal phase velocity is $v_{long}^1 \equiv dx^1/dt = \omega/k_1$. We obtain
\begin{equation}
v_{long}^1= \frac{1}{R}\sqrt{1+\frac{{\bf B_0}^2}{2\rho_0}}.
\end{equation}
\label{48}
The longitudinal  phase speed is thus
\begin{equation}
v_{long}= \sqrt{g_{11}}v_{long}^1=  \sqrt{  1+\frac{{\bf B_0}^2}{2\rho_0}}.
\end{equation}
\label{49}
This expression is independent of the scale factor $R$. It has the same structure as the standard expression $v_{long}=\sqrt{s^2+v_A^2}$ for the longitudinal magnetosonic wave in nongravitational MHD \cite{jackson75}, $s$ denoting the sound velocity, if we define the Alfv\'{e}n velocity as
\begin{equation}
   {\bf v_A}={\bf B_0}/\sqrt{2\rho_0}.
\end{equation}
\label{50}
In the present case, $s=1$.

The following property of the dispersion relation (47) is however striking: the expression is nondispersive, which implies that the {\it group velocity}  $ \partial {\omega}/\partial {k} $ in this case becomes {\it superluminal}. This is physically non-acceptable. We thus have to conclude that the Zel'dovich fluid model, being a maximally rigid model, does not permit the  longitudinal magnetosonic wave to propagate as a physical wave, transporting energy. This result is not so unexpected, after all. The Zel'dovich model is very simple, and has in practice to be modified so as to take into account dispersive effects. 

Next, let ${\bf B_0}$ be {\it parallel } to ${\bf k}$. There are two types of wave motion possible in this case. There is an ordinary longitudinal wave ($u_{02}=u_{03}=0, ~{\bf b_0}=0$) with contravariant phase velocity component equal to $1/R$ and phase speed accordingly equal to 1. There is also a {\it transverse} wave ($u_{01}=0$), where ${\bf b_0}$ is antiparallel (if $ {\bf k \cdot B_0 }  > 0 $) to the transverse velocity ${\bf u_0}$:
\begin{equation}
 {\bf b_0} = -\frac{R^3}{\omega}(\bf{k \cdot B_0})\bf{u_0}.
\end{equation}
\label{51}
The dispersion relation becomes in this case
\begin{equation}
\omega^2=\frac{1}{2\rho_0}({\bf k\cdot B_0})^2,
\end{equation}
\label{52}
so that the contravariant component of the transverse phase velocity becomes
\begin{equation}
v_{tr}^1= \frac{1}{\sqrt{2\rho_0}}B_0^1.
\end{equation}
\label{53}
The speed of the transverse wave is thus 
\begin{equation}
v_{tr}=\frac{B_0}{\sqrt{2\rho_0}};
\end{equation}
\label{54}
again an expression that is independent of the scale factor. We see that Eq. (54) actually agrees with the expression (50) for the Alfv\'{e}n velocity; as noted above, this correspondence is necessary in order to preserve the same formal relationship    $v_{long}=\sqrt{s^2+v_A^2}$ as in nonrelativistic theory.

To obtain a feeling of the order of magnitudes involved, let us estimate the value of $\rho_0$ to be inserted in the expression (50) for the Alfv\'{e}n velocity. Recall that $\rho_0$ is the density of matter and radiation; it does not include the vacuum energy $\rho_{vac}=\Lambda/8\pi G$. We estimate $\rho_0$ by considering the instant $t=t_1=1.4 \times 10^{-33}$ s  just after the termination of the inflationary era, when the Universe leaves the de Sitter phase and returns to the standard FRW radiation dominated form.  When  $k=0$, the density is equal to the critical density
\begin{equation}
\rho_{cr}=\frac{3}{8\pi}\frac{H^2}{G}.
\end{equation}
\label{55}
(It is noteworthy that this critical density follows from Einstein's equations directly, for any metric that is expressible in the form (5); the expression does not rely upon a specific state equation.) Thus $\rho_0=\rho_{cr}$ at $t=t_1$. Taking the Hubble factor $H \equiv \dot{R}/R$ during the inflationary era to be $H=5\times 10^{34}~ {\rm s^{-1}}$  \cite{gron86}, we obtain from Eq. (55) $\rho_{cr}=4.5 \times 10^{75}~ {\rm g/cm^3} $. This is an enormous mass density. The magnitude of the Alfv\'{e}n velocity (50) becomes accordingly extremely small:
\begin{equation}
v_A= 3 \times 10^{-37} B_0~{\rm (tesla)~~ m/s} = 3 \times 10^{-31} B_0 ~{\rm (gauss)~~ cm/s}.
\end{equation}
\label{56}
The transverse Alfv\'{e}n wave (a purely magnetohydrodynamic phenomenon) thus does not propagate appreciably under these extraordinary circumstances, unless the value of ${\bf B_0}$ is extremely high. And, as far as we know, this is not the case in the early Universe.

\section{Summary and final remarks}

We may summarize as follows:

(1)  Our basic assumption is the presence of a Zel'dovich fluid, satisfying $p=\rho$, in a de Sitter universe whose metric is given by Eq. (5). The scale factor is given by Eq. (11). There is a uniform background magnetic field ${\bf B_0}$ present. The magnetic field energy is taken to be not so high that it contributes appreciably as a source term in Einstein's equations. The electrical conductivity is assumed to be infinite. The fluid velocities in the comoving reference system are taken to be nonrelativistic. The most actual example of application is the inflationary era, ranging from $t_0=10^{-35}$ s to $t_1=10^{-33}$ s after big bang.

(2)  For a de Sitter universe, energy conservation considerations lead to the property $\rho R^6 $= const., instead of the conventional property $\rho R^4 $=const. holding for a radiation dominated FRW universe. Considering plane, linear, MHD waves on the uniform de Sitter background, we find that the longitudinal magnetosonic wave (${\bf B_0 }\perp {\bf k}$) implies a superluminal group velocity and accordingly does not correspond to a physical, energy-transporting, wave. Of the two remaining fundamental modes $({\bf B_0} \parallel {\bf k})$, the ordinary longitudinal wave (with vanishing perturbed magnetic field) can exist as a luminal wave, and there is a third kind of wave of  magnetohydrodynamic type propagating with the characteristic Alfv\'{e}n velocity given by Eq. (50).

(3)  Estimating the value of $\rho_0$ to be equal to the critical mass density of a flat FRW universe, Eq. (55), we find that the Alfv\'{e}n velocity becomes extremely small, Eq. (56). The magnetohydrodynamic wave in practice does not propagate; it is 'frozen' in the fluid. 

(4)  The following remark ought to be made, concerning the physical meaning of the present theory. Our adopted state equation, $p=\rho$, refers to the content of {\it matter}, and {\it radiation}, in the early Universe. One may ask: is not the early Universe after all dominated by {\it vacuum} effects, corresponding to the state equation $p_{vac}=-\rho_{vac}$?  The point here is that both these state equations can exist at the same time; they refer to two different subsystems. The properties of the vacuum fluid are described entirely by the $\Lambda$ term in the modified energy-momentum tensor $\tilde{T}_{\mu\nu}$ in Eq. (2); as mentioned earlier it corresponds to a vacuum energy $\rho_{vac}=\Lambda/8\pi G$ and a vacuum pressure $p_{vac}=-\Lambda/8\pi G$. In addition to this, the matter/radiation fluid forms a separate subsystem. 

The point that we wish to stress is: although the present theory in itself does not make a definite prediction about the relative energy fraction residing in the matter component, some amount of matter has always to be present, in order to support the magnetic field. In a pure de Sitter universe consisting of  expanding vacuum (cosmological constant only) there would be no charge carriers from which a magnetic field could be generated. Even if we imagine that there was for some reason a magnetic field present at some instant of time, this field would disappear quickly due to the expansion. It gives no physical meaning to consider MHD in an empty, purely vacuum, universe.

(5)  It has recently come to our attention that Subramanian and Barrow have published a detailed study \cite{subramanian98a} of consequences of MHD in the early FRW Universe. They discuss various aspects of the damping problem and also the possibility of observational detection of the magnetic fluctuations. Cf. also the related Ref. \cite{subramanian98b}.

\bigskip
\bigskip

{\bf Acknowledgments}
\bigskip

We thank Professors {\O}yvind Gr{\o}n and John D. Barrow for valuable discussions and information.

\renewcommand{\theequation}{\mbox{\Alph{section}\arabic{equation}}}
\appendix
\setcounter{equation}{0}

\section{Remarks on the FRW case}

We will give a brief account of the analogous theory for a radiation dominated flat FRW universe, both to put our above results into perpective, and also because our results for the dispersion equations for the modes differ partly from those obtained in \cite{brandenburg96}.

The line element still has the form (5), but the scale factor $R$ now becomes {\cite{gron86}
\begin{equation}
\frac{R(t)}{R_0}=\left( \frac{32 \pi}{3}\rho_0 \right)^{1/4}t^{1/2}.
\end{equation}
\label{A1}
The equation of state for the radiation dominated fluid is 
\begin{equation}
p=\frac{1}{3}\rho.
\end{equation}
\label{A2}
The governing equations are the $\mu=0$ equation (25), the $\mu=i$ equation (26), and the Maxwell equation (34), as before. Taking Eq. (A2) into account, we obtain for the $\mu=0$ equation
\begin{equation}
\frac{1}{R^4}\partial_0(R^4\rho)+\frac{4}{3}{\bf \nabla \cdot (\rho {\bf u})}={\bf E \cdot J},
\end{equation}
\label{A3}
whereas the $\mu=i$ equation becomes
\begin{equation}
\frac{4}{3R^3}\partial_0(R^3\rho u_i)+\frac{1}{3}\partial_i \rho= (B^k\partial_k)B_i-\frac{1}{2}\partial_i B^2.
\end{equation}
Considering plane, linear, MHD waves on a uniform background in this universe we now have, instead of Eq. (35),
\begin{equation}
\rho_0(t) \propto \frac{1}{R^4(t)}, ~~~~~B_0^i(t) \propto \frac{1}{R^3(t)}.
\end{equation}
\label{A5}
We expand the wave quantities analogously to Eq. (41) (for simplicity keeping the same symbols as previously for the constants):
\begin{equation}
u^i=u_0^i e^{i\Theta}, ~~~\delta\rho=\frac{C}{R^4}e^{i\Theta},~~~\delta B^i=\frac{b_0^i}{R^3}e^{i\Theta}.
\end{equation}
\label{A6}
We obtain the conservation equations
\begin{equation}
\frac{4}{3}\rho_0 R^4 {\bf k\cdot u_0}=C\omega,
\end{equation}
\label{A7}
\begin{equation}
\frac{4}{3}\omega\rho_0 R^3 u_{0i}-\frac{C}{3R}k_i=({\bf b_0 \cdot B_0})k_i-({\bf k \cdot B_0})b_{0i},
\end{equation}
\label{A8}
which together with Maxwell's equation (46) determine the dispersion relations. We distinguish between the same cases as above:

If ${\bf B_0}$ is {\it perpendicular} to ${\bf k}$, then there is a longitudinal wave ($ u_{02}=u_{03}=0$) with dispersion relation
\begin{equation}
\omega^2= \left( \frac{1}{3}+\frac{3{\bf B_0}^2}{4\rho_0} \right){\bf k}^2.
\end{equation}
\label{A9}
This equation, contrary to Eq. (47), describes a physical disturbance, since both phase velocity and group velocity are subluminal.

If ${\bf B_0}$ is {\it parallel} to ${\bf k}$, there is an ordinary longitudinal nondispersive luminal wave, as before. Moreover there is a transverse wave whose dispersion relation is
\begin{equation}
\omega^2=\frac{3}{4\rho_0}({\bf k\cdot B_0})^2.
\end{equation}
\label{A10}
Comparison between Eqs. (47) and (A9), and between Eqs. (52) and (A10), shows how the de Sitter-Zel'dovich case differs from the radiation dominated FRW case. Whereas the longitudinal wave (A9) agrees with Eq. (31) in \cite{brandenburg96}, there appears to be a deviation in case of  the transverse wave, Eq. (A10). It is also noteworthy that Eq. (A9) is in agreement with the relationship $v_{long}=\sqrt{s^2+v_A^2}$, where now $s=\sqrt{1/3}$ and ${\bf v_A}=\sqrt{3/4\rho_0}\,{\bf B_0}$ is the FRW Alfv\'{e}n velocity corresponding to Eq. (A10).

\end{document}